# Microstructure, mechanical properties and corrosion of friction stir welded 6061 Aluminum Alloy

Zhitong Chen[*], Shengxi Li, Lloyd H. Hihara[*]

Hawaii Corrosion Laboratory, Department of Mechanical Engineering, University of Hawaii at Manoa, Honolulu, HI 96822, USA

*Abstract*

The microstructure, mechanical properties, and corrosion behavior of friction stir welded (FSW) AA6061 aluminum alloys were investigated. Dynamic recrystallized structures were observed and grain sizes of nugget zone (NZ), thermomechanically-affected zone (TMAZ), heat-affected zone (HAZ), and base material (BM) were different. Hardness test indicated that the minimum and maximum hardness values wereobtained in the HAZ and BM, respectively. Tensile results showed that fracture occurred in the relatively weak regions in between TMAZ and HAZ. Polarization tests illustrated that the FSW process improved the corrosion resistance of AA6061-AA6061 and the HAZ had better corrosion resistance than other regions. Raman characterizations revealed that aluminum hydroxide was the main corrosion product formed on Al after immersion experiments. Intergranular attack was observed in the NZ and downside by scanning electron microscopy.

*Keywords*

Corrosion, Friction Stir Weld, Aluminum Alloy 6061, Mechanical Properties

---

[*] Corresponding Author:
E–mail address: Jeadonchen@gmail.com, hihara@hawaii.edu

## 1 Introduction

Friction stir welding (FSW), a new solid state joining technique, was invented by The Welding Institute (TWI) in 1991 [1]. FSW is becoming more popular for joining a wide range of aluminum alloys for numerous applications. One advantage of FSW is that there is far lower heat input during the process compared with conventional welding methods such as TIG or MIG. Therefore, this solid state process results into minimal microstructural changes and better mechanical properties than conventional welding [2-4]. The FSW process generates three distinct microstructural zones: the nugget zone (NZ), the thermomechanically-affected zone (TMAZ) and the heat-affected zone (HAZ) [5-9]. The HAZ is only affected by heat, without plastic deformation. The TMAZ adjacent to the nugget is plastically deformed and heated. The nugget is affected by the highest temperature and the highest plastic deformation, which generally consists of fine equiaxed grains due to the full dynamic recrystallization.

The microstructural changes induced by the plastic deformation and the frictional heat during FSW process have been extensively studied [7-15]. A relationship between microstructure and microhardness of each FSW weld zone has been discovered [2, 4]. Changes in microhardness along the FSW joint are directly related to the precipitation state. Research on the corrosion behavior of FSW joint has shown that FSW regions are more susceptible to localized corrosion than the parent metal in 2xxx, 7xxx and 5xxx series Al alloys [16-24]. For AA2024–T3, corrosion attack in the nugget has been found [16]. For AA7050–T7651, attack was found at the interface between the nugget and the partially recrystallized zone (TMAZ) [15]. For AA7075–T651 and AA2024–T351, studies have shown attack to be predominantly in the HAZ [17, 24]. It has also been found that the weld region can show no worse corrosion susceptibility than the base metal in AA2024–T3 and AA2195 welds [21], and even in some cases, improved corrosion resistance compared with the base metal, for example in AA5083, AA2024–T3, and AA2219 [19, 22, 23]. Among these works, only few works exhibited a variation of electrochemical surface reactivity between the various zones of a FSW weld. There remains a limited understanding of the relationship between the microstructure and corrosion of FSW in Al alloys, especially for AA6061. Moreover, there has not been any systematic study reporting the relationship between microstructure and corrosion behavior of AA6061.

In this work, in order to determine the local corrosion behavior along the weld, an experimental investigation with localized immersion, electrochemical and mechanical measurements were carried out combined with SEM and Raman. A relationship between microstructure, microhardness and corrosion sensitivity of a friction stir welded 6061 aluminum alloy was established.

*2 Experimental procedure*

AA 6061 (0.4% Si, 0.7% Fe, 0.4% Cu, 0.15% Mn, 1.2% Mg, 0.35% Cr, 0.25% Zn, 0.15% Ti, balance Al) plates were friction stir welded vertical to the rolling direction with a travel speed, a rotational speed and a shoulder diameter of 20 mm/min, 1000 rpm and 25 mm. The friction stir pin had a diameter of 8 mm and height of 6.35 mm. A simultaneous rotation and translation motion of the FSW tool generates the formation of an asymmetric weld. When the tool rotates in the direction of its translation, it refers to the advancing side (AS). When rotation and translation of the tool are in the opposite direction, it refers to the retreating side (RS).

The grain structure of the welds at different regions was revealed by etching in Keller's reagent and observed by optical microscopy. Vickers microhardness testing (Wilson Rockwell, R5000) was performed across the cross section of the welds at a distance of 2 mm from the top surface using a 50 g load. The tensile tests were performed at a crosshead speed of 3 mm/min using an Instron-5500R testing machine. Tensile specimens were machined from the NZ in two directions from the weld: parallel (longitudinal) and normal (transverse). The configuration of the tensile specimens is shown in Fig. 1. An axial extensometer with 25 mm gage length was attached to the specimens at the gauge section. The strain analysis of each specimen was made by an ASAME automatic strain measuring system. The tensile properties of the joints were evaluated using three tensile specimens cut from the same joint.

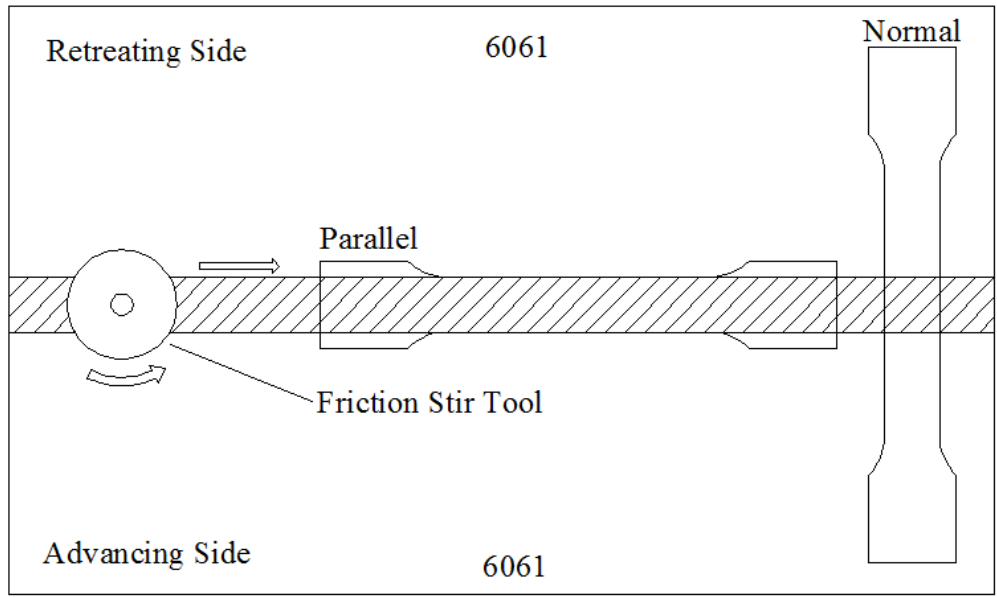

Fig.1 Schematic of the locations from where tensile test specimens were cut from (P-AA6061 (Longitudinal) and N-AA6061 (Transverse))

Polarization experiments were conducted using samples cut from different zones, i.e.,NZ, TMAZ, HAZ and BM. The samples were mounted in epoxy resin, polished to a 0.05 μm mirror-finish, and immersed in high purity water (18MΩ·cm) prior to polarization experiments in 3.15 wt% NaCl solutions at 30°C. The solution was deaerated with high-purity nitrogen (> 99.999%). Potentiodynamic polarization experiments were conducted with a potentiostat (PARSTAT 2273, Advanced Electrochemical System). The working electrodes were kept in the open-circuit condition for 1 hour prior to conducting the potentiodynamic scan at a rate of 1 mV/s. A saturated calomel electrode (potassium chloride) was used as the reference electrode and a platinum mesh was used as the counter electrode. To minimize contamination of the solution, the reference electrode was kept in a separate cell connected via a Luggin probe. Anodic and catholic sweeps were measured separately on freshly-prepared surfaces. Unless specified otherwise, polarization experiments for each typical zone were performed using at least three samples to verify reproducibility. To generate polarization diagrams, the mean values of the logarithm of the current density were plotted as a function of potential.

In immersion tests, specimens of FSW (70 mm × 25 mm × 4 mm) were cleaned and dried. The coupons were then degreased in acetone, ultrasonically cleaned in deionized water, dried and weighted prior to the experiments. Three coupons of each material, mounted to an acrylic holder, were placed in a 250 ml beaker. A total of 24 beakers were kept at a uniform temperature of 30 ºC by immersing the beaker in a heated and water-circulated aquarium. Approximately 200 ml of 3.15 wt% NaCl, 0.5M $Na_2SO_4$, and ASTM seawater solution was poured into the beaker so that the coupons were completely immersed in the solution. Beakers were partially covered to minimize evaporation and to maintain an aerated condition. After 90 and 120 days of immersion, the coupons were retrieved and analyzed by SEM and Raman spectroscopy. The corroded coupons were cleaned in a solution of phosphoric acid ($H_3PO_4$) and chromium trioxide ($CrO_3$) at 90 ºC for 10 minutes as described in ASTM G01-03 and their weight loss were noted after drying.

## 3 Results and Discussion

### 3.1 Microstructure

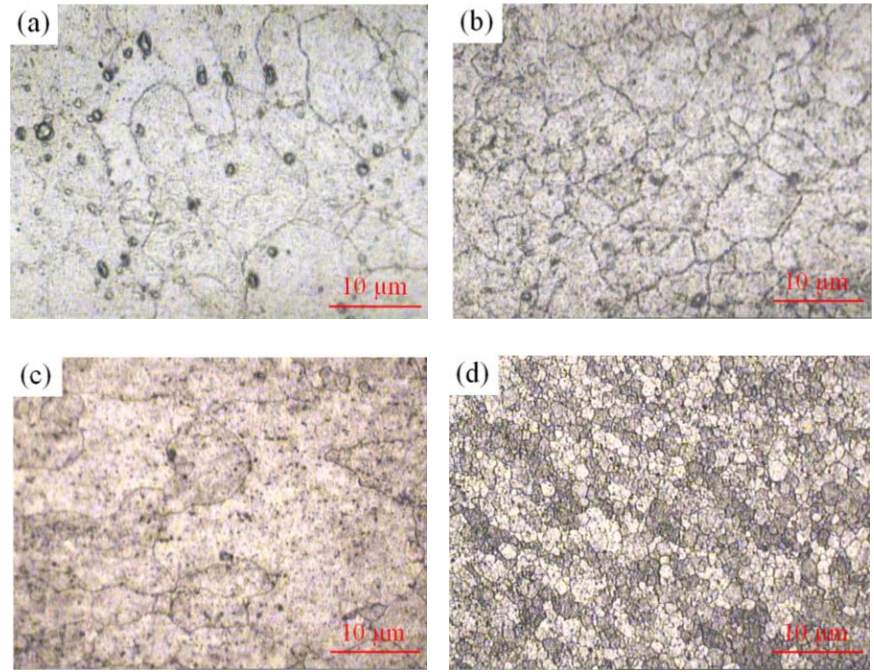

Fig. 2 Typical microstructures at different regions of FSW AA6061-AA6061 after etching with Keller's reagent: (a) HAZ, (b) TMAZ, (c) BM, and (d) NZ

Fig. 2 shows typical grain structures of different regions in FSW 6061-6061. The NZ consists of fine equiaxed grains due to dynamic recrystallization [25, 26]. The grains in NZ are much smaller than those in other regions. The average grain size in the four zones in follows the order of BM > HAZ > TMAZ > NZ. In the TMAZ which is adjacent to the NZ, the strain and the temperature were lower than in the NZ and the effect of welding on the microstructure was correspondingly smaller. Unlike NZ, the microstructure was recognizably that of the parent material, although significantly deformed and rotated. The grain size of the HAZ was similar to that of the BM. The HAZ was common to all welding processes subjected to a thermal cycle, but it was not deformed during welding.

*3.2 Mechanical properties*

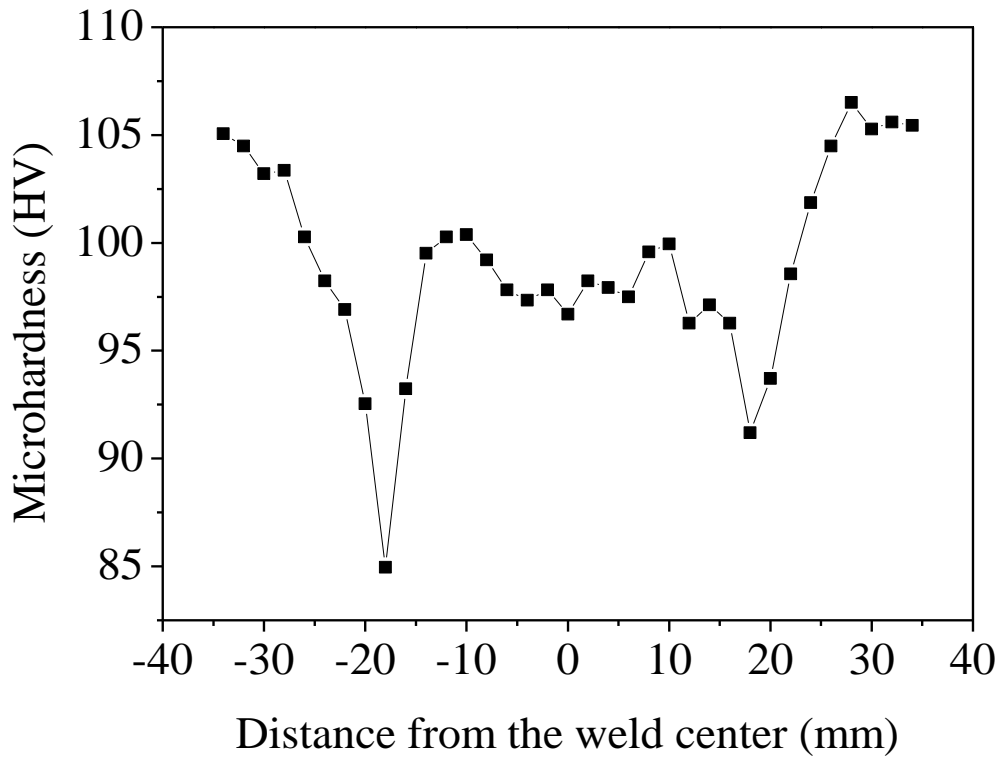

Fig. 3 Microhardness distribution across the top surface of FSW AA6061-AA6061 measured with a 2 mm step

Fig. 3 shows the typical microhardness distribution across the top surface of FSW AA6061-AA6061. The hardness curve is asymmetrical with respect to the weld centerline because the plastic flow field in the two sides of the weld center is not uniform [27, 28]. The larger distorted grains and distortion energy causes the strain-hardness to increase, resulting in the asymmetrical microhardness distribution. The minimum hardness of 85.0 HV was obtained in the HAZ region, suggesting that the tensile specimens are prone to fracture in this zone. The maximum value was present in the BM. The hardness of the TMAZ was higher than that of the NZ. However, according to the Hall-Petch relationship [29], the hardness value in the NZ should be higher than other zones because of its fine equiaxed grain structure. . The reason for the NZ having lower hardness values than those in TMAZ is possibly due to the thermal history of the FSW AA6061-AA6061 which creates a softened region around the weld center. The possible reasons for softening are that the thermal cycle of FSW causes strengthening precipitates to coarsen or dissolve. To analyze the hardness, the Orowan mechanism, which is based on the impediment of dislocation movement around small precipitates inside grains, may over shadow the Hall-Petch mechanism, which is based on grain size [29, 30].

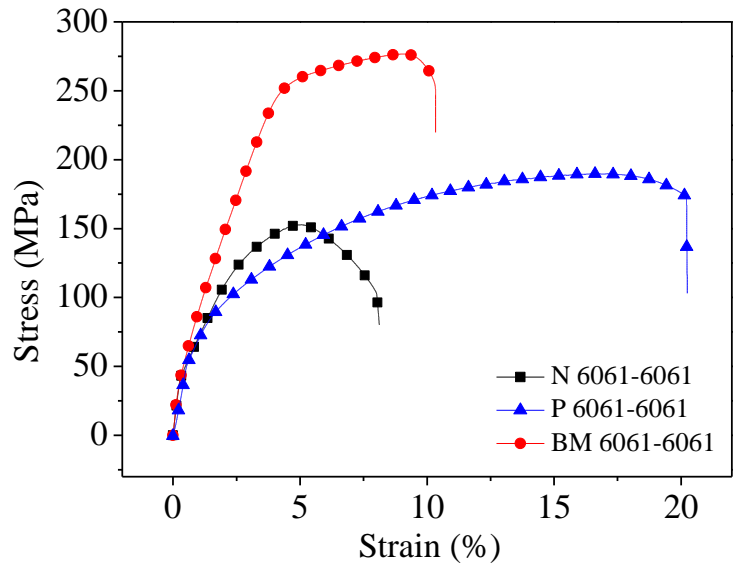

Fig. 4 Tensile properties of BM and two FSW specimens cut from different orientations (P: parallel, N: normal, see Fig. 1).

As shown in Fig. 4, the elongation, yield strength, and tensile strength of the BM AA6061 were 12.1%, 242 MPa, and 276 MPa, respectively. By comparison, the two FSW specimens showed a significant decrease in both tensile and yield strength. Fig. 4 also show that the the ductility for tensile specimen P AA6061 significantly increased compared to that of the BM while that of specimen N-AA6061 decreased. As indicated in Fig.1, the tensile specimen P-AA6061 contained only recrystallized grains from the NZ. From the hardness results, we know that the hardness values of the NZ were lower than that of the BM, possibly explaining why the longitudinal tensile specimen P-AA6061 exhibited both tensile and yield strength values. The transverse tensile specimens contained all four zones (i.e., BM, HAZ, TMAZ and NZ). The observed ductility was measured as average strain over the gage length, including the various zones that have different resistances to deformation due to differences in grain size and precipitate distribution. When a tensile load was applied to the joint, failure occurred in the weakest regions of the joint [31], which is the HAZ in this work.

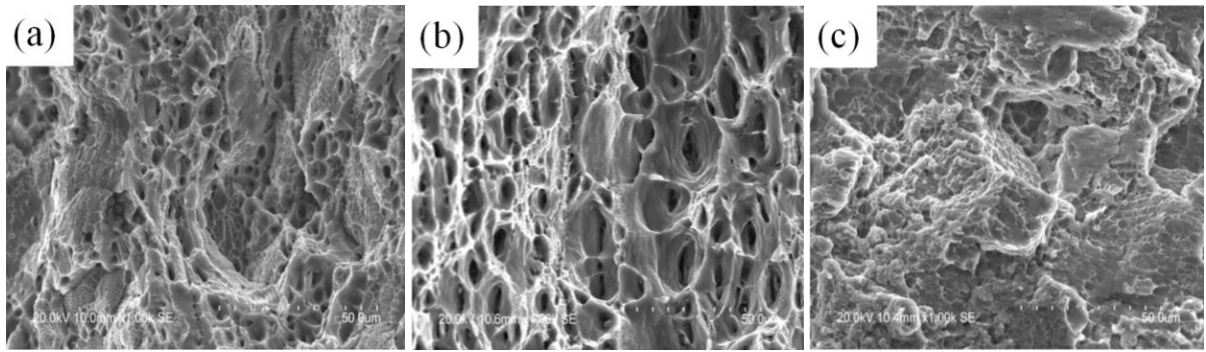

Fig. 5 SEM fractographs of BM and two directions (Transverse and Longitudinal) FSW specimens: (a) BM AA6061-AA6061, (b) T AA6061-AA6061 (Failure occurred in HAZ), and (c) L AA6061-AA6061 (Failure occurred in Weld Zone)

Fig. 5 shows the fracture surfaces of tensile specimens characterized by SEM. The tensile fractures of L AA6061-AA6061 presented a 45° angle shear fractures along the tensile axis, while BM AA6061-AA6061 and T AA6061-AA6061 (failure occurred in HAZ) presented a 90°. The fracture surfaces of T AA6061-AA6061 and L AA6061-AA6061 showed obvious necking/plastic deformation except the BM AA6061. The fractographs reveal dimple fracture patterns with teared edges full of micropores. The dimples were of various sizes and shapes. . Compared to L AA6061-AA6061, T AA6061-AA6061 had deeper dimples and thinner teared edges. Therefore, the L AA6061-AA6061 exhibited much better mechanical properties than did the BM AA6061. Thus, the longitudinal direction FSW specimens have much better mechanical properties than do the transverse direction specimens.

### 3.3 Electrochemical measurements

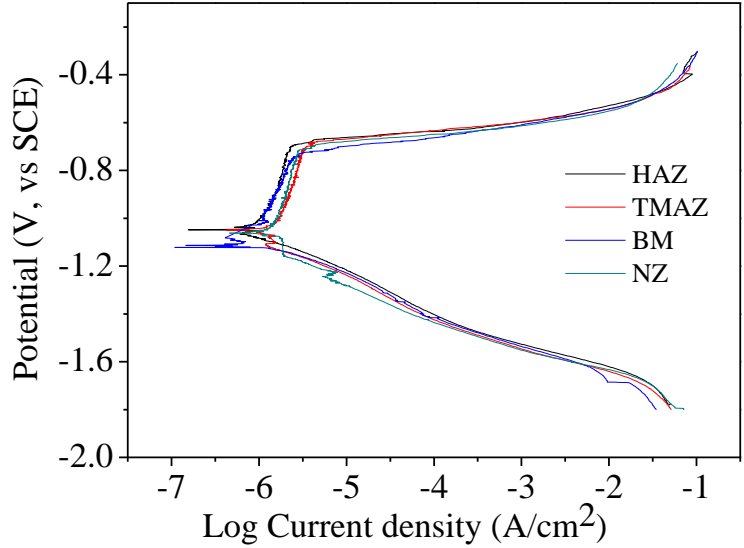

Fig. 6 Polarization curves of different weld zones of FSW AA6061-AA6061 in deaerated 3.15 wt% NaCl solutions

Table 1 $E_{corr}$, $I_{corr}$ and $E_{pit}$ values of the different weld zones of FSW AA6061-AA6061 in deaerated 3.15 wt% NaCl solutions and standard deviation (S.D.) of $I_{corr}$

| Weld | $E_{corr}$ (mV, vs SCE) | $E_{pit}$ (mV, vs SCE) | $I_{corr}$ (μA/cm$^2$) | S.D. of $I_{corr}$ |
|---|---|---|---|---|
| HAZ | -1048 | -692 | 0.76 | 0.03 |
| TMAZ | -1049 | -685 | 1.54 | 0.27 |
| BM | -1114 | -724 | 2.44 | 0.71 |
| NZ | -1063 | -719 | 2.29 | 0.38 |

Fig. 6 shows polarization curves of different weld zones of FSW AA6061-AA6061 in deaerated 3.15 wt% NaCl solutions. All four samples showed passive region with current density in the range of $10^{-6}$ and $10^{-5.5}$ A/cm². Compared to the BM, the three samples from weld zones had higher $E_{corr}$ and $E_{pit}$ values and lower $I_{corr}$ values (Table 1), indicating that the corrosion resistance was improved by FSW. The order of corrosion potential for samples from different zones was as follows: HAZ (-1048 mV$_{SCE}$) > TMAZ (-1049 mV$_{SCE}$) > NZ (-1063 mV$_{SCE}$) > BM (-1114 mV$_{SCE}$). For the $E_{pit}$, the order was as follows: TMAZ (-685 mV$_{SCE}$) > HAZ (-692 mV$_{SCE}$) > NZ (-719 mV$_{SCE}$) > BM (-724 mV$_{SCE}$). For $I_{corr}$, the values from highest to lowest were as follows: BM (2.44 µA/cm²) > NZ (2.29 µA/cm²) > TMAZ (1.54 µA/cm²) > HAZ (0.76 µA/cm²). The dissolution of finer precipitates occurred both in the NZ and the downside. The coarsening of precipitates in the TMAZ and HAZ could be the reason for the improved corrosion resistance [32].

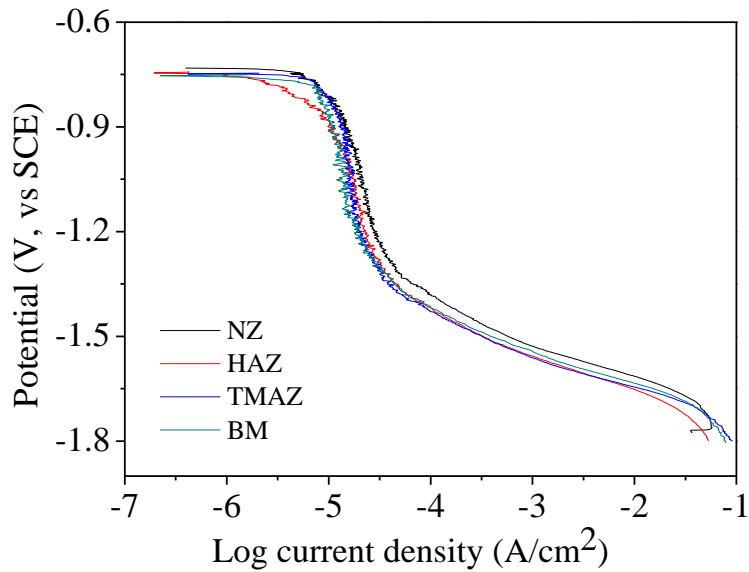

Fig. 7 Cathodic polarization curves of different weld zones of FSW AA6061-AA6061 in aerated 3.15 wt% NaCl solutions

Table 2 *Ecorr* and *Icorr* values of the different weld zones of FSW AA6061-AA6061 in aerated 3.15 wt% NaCl solutions and standard deviation (S.D.) of *Icorr*

| Weld | $E_{corr}$ (mV, vs SCE) | $I_{corr}$ (μA/cm$^2$) | S.D. of $I_{corr}$ |
|---|---|---|---|
| HAZ | -742 | 2.87 | 0.82 |
| TMAZ | -746 | 6.41 | 0.07 |
| BM | -750 | 8.57 | 1.24 |
| NZ | -731 | 5.02 | 1.56 |

Fig. 7 shows cathodic polarization diagram of various weld zones of FSW AA6061-AA6061 in aerated 3.15 wt% NaCl solutions. In aerated 3.15 wt% NaCl, different weld zones of FSW AA6061-AA6061 showed diffusion-limited oxygen reduction, which was different from FSW AA5086-AA5086 (references?). The $E_{corr}$ and $I_{corr}$ values of the various weld zones of FSW AA6061-AA6061 in aerated 3.15 wt% NaCl solutions are shown in Table 2. The $E_{corr}$ values of HAZ (-742 mV$_{SCE}$), TMAZ (-746 mV$_{SCE}$), and NZ (-731 mV$_{SCE}$) shifted to more positive values from that of the BM (-750 mV$_{SCE}$). The $E_{corr}$ values were also different from those for FSW AA5086-AA5086 (references?). The $I_{corr}$ values decreased from that of the BM (8.57 μA/cm$^2$). These results show that FSW improved the corrosion resistance of welded AA6061-AA6061.

### *3.4 Corrosion attack*

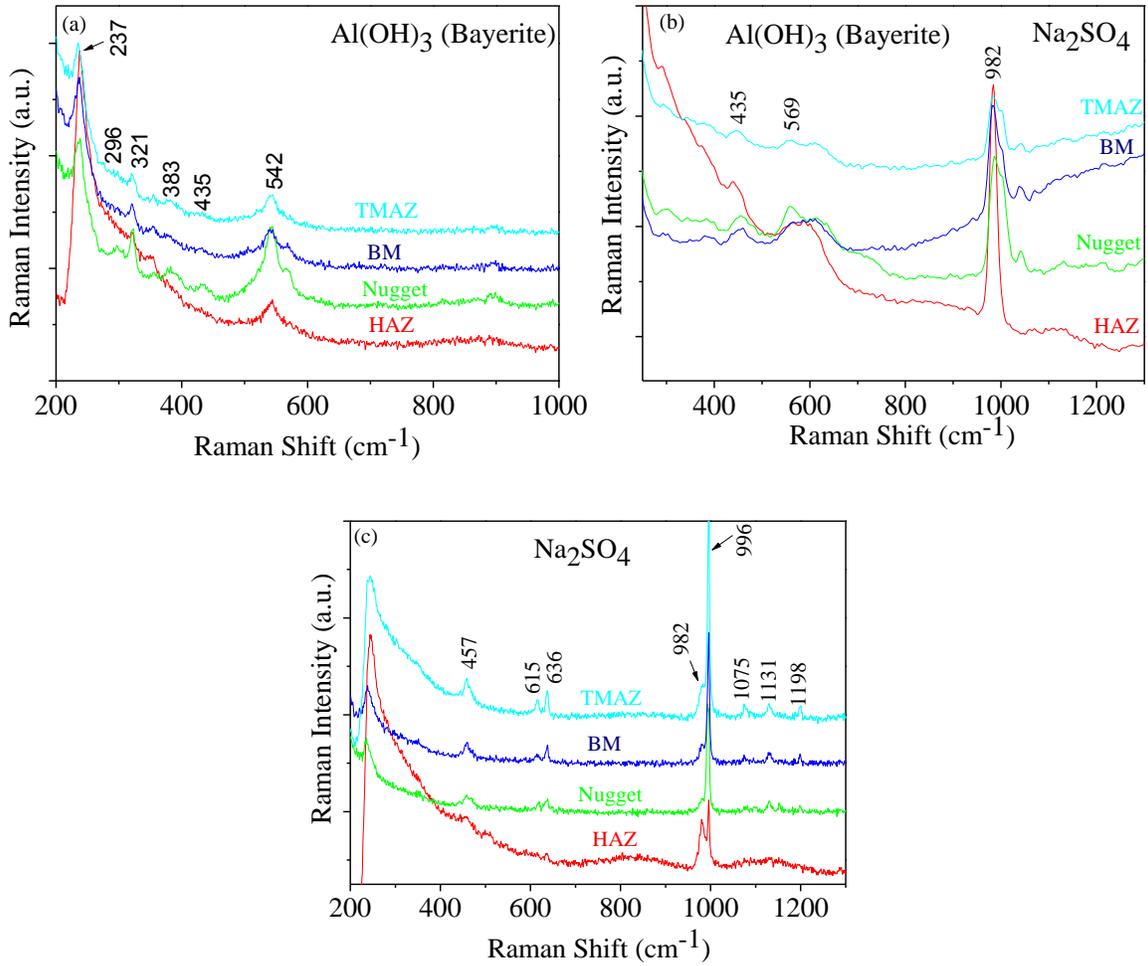

Fig. 8 Raman spectra of the corrosion products formed in distinct zones of FSW AA6061-AA6061 after 90 days immersion in: (a) 3.15 wt% NaCl, (b) ASTM seawater, and (c) 0.5 M Na$_2$SO$_4$ solution.

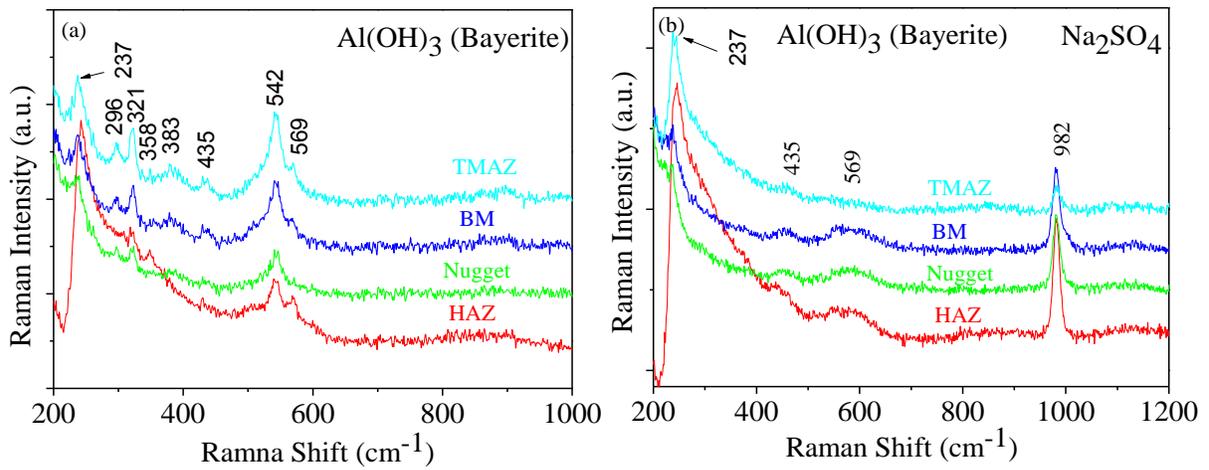

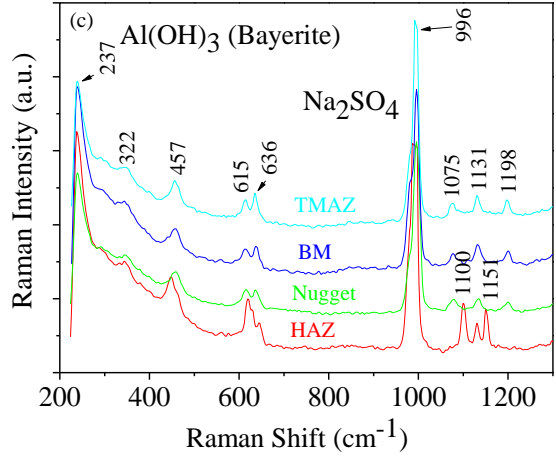

Fig. 9 Raman spectra of the corrosion products formed in distinct zones of FSW AA6061-AA6061 after 120 days immersion in: (a) 3.15 wt% NaCl, (b) ASTM seawater, and (c) 0.5 M $Na_2SO_4$ solution

Figs. 8 and 9 show Raman spectra of the corrosion products formed in distinct zones of FSW AA6061-AA6061 after 90 and 120 days immersion in 3.15 wt% NaCl, ASTM seawater, and 0.5 M $Na_2SO_4$ solution, respectively. $Al(OH)_3$ was identified as the corrosion products formed on samples (from all FSW regions) that were immersed in 3.15 wt% NaCl for both 90 and 120 days (Fig. 8a and Fig. 9a). The Raman intensities of characteristic bands increased as the immersion time increased from 90 to 120 days. Raman spectra from samples immersed in ASTM seawater (Fig. 8b and Fig. 9b) show strong bands of $Na_2SO_4$ (982 $cm^{-1}$) and relatively weak bands of $Al(OH)_3$ (435 and 459 $cm^{-1}$). A characteristic band (237 $cm^{-1}$) of was found only in Fig. 9b. In Fig. 8(c), the characteristic bands of $Na_2SO_4$ (457, 615, 636, 982, 996, 1075, 1131, and 1151 $cm^{-1}$) were observed on the TMAZ, BM, and NZ, while only characteristic bands of $Na_2SO_4$ (982 and 996 $cm^{-1}$) were observed in the HAZ. As shown in Fig. 8 (c), the bands that were not marked were ghost band [33-35]. Raman bands of $Al(OH)_3$ for FSW AA6061-AA6061 after immersion 90 days in 0.5 M $Na_2SO_4$ solution were missing because the amount of corrosion products may have been insufficient to generate Raman peaks. As shown in Fig. 9(c), characteristic bands of

Al(OH)$_3$ (237, 322, and 457 cm$^{-1}$) and Na$_2$SO$_4$ (615, 636, 996, 1100, 1131, and 1151 cm$^{-1}$) were observed in four regions, a finding that tells us that the amount of Al(OH)$_3$ increased for FSW 6061-6061 coupons immersed in 0.5 M Na$_2$SO$_4$ solution from 90 to 120 days. There were also characteristic bands (1075 and 1198 cm$^{-1}$) that cannot be identified.

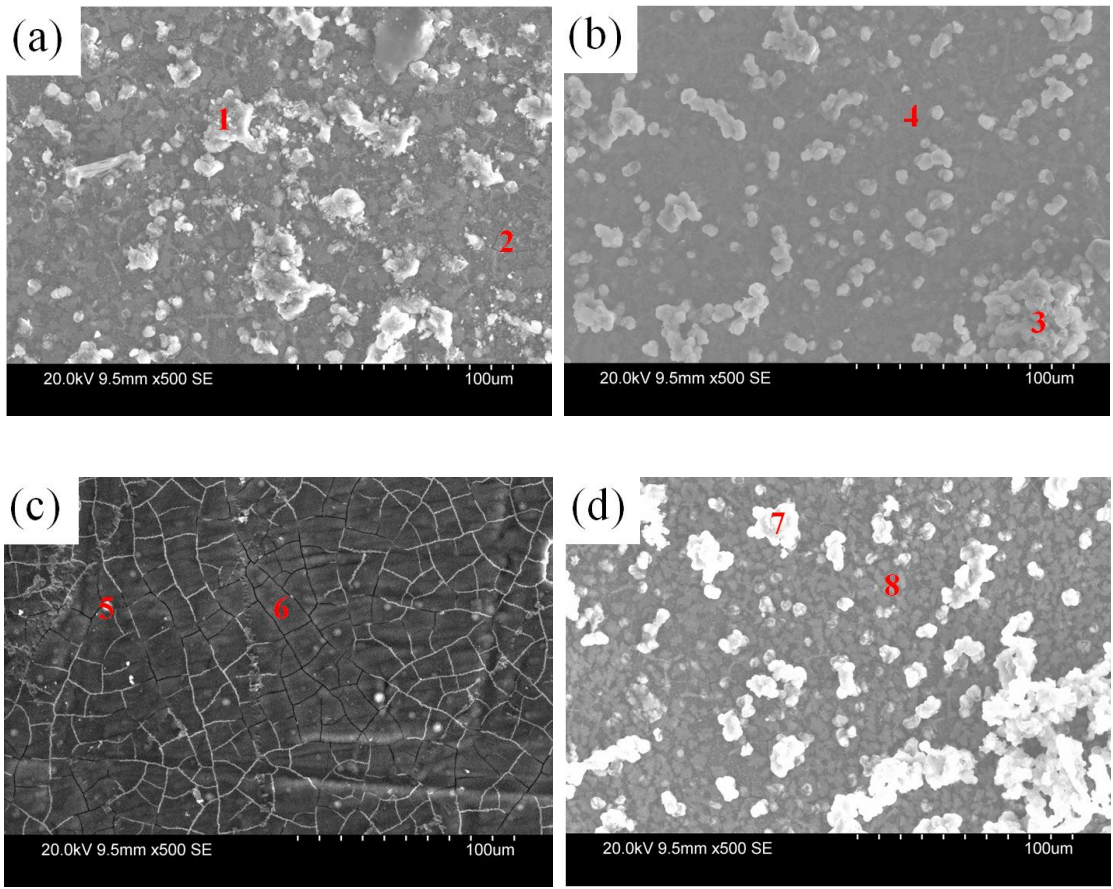

Fig. 10 SEM images of FSW AA6061-AA6061 after 90 days of immersion in 0.5 M Na$_2$SO$_4$: (a) NZ, (b) TMAZ, (c) HAZ, and (d) BM

Table 3 EDXA quantification results of eight points in Fig. 10

| Element [at. %]   | C     | O     | Na    | Al    | S    |
|-------------------|-------|-------|-------|-------|------|
| Spectrum 1 (NZ)   | 8.99  | 59.89 | 9.70  | 14.47 | 6.95 |
| Spectrum 2 (NZ)   | 0.00  | 51.05 | 3.59  | 41.85 | 3.51 |
| Spectrum 3 (TMAZ) | 10.21 | 66.62 | 5.56  | 11.93 | 5.68 |
| Spectrum 4 (TMAZ) | 0.00  | 59.40 | 5.50  | 31.49 | 3.61 |
| Spectrum 5 (HAZ)  | 16.23 | 61.74 | 5.48  | 13.73 | 2.82 |
| Spectrum 6 (HAZ)  | 8.85  | 56.92 | 2.40  | 29.77 | 2.06 |
| Spectrum 7 (BM)   | 6.72  | 65.79 | 11.61 | 11.33 | 4.55 |
| Spectrum 8 (BM)   | 5.28  | 50.85 | 4.39  | 35.66 | 3.81 |

The SEM images of five regions of FSW 6061-6061 after 90 days immersion in 0.5 M $Na_2SO_4$ solution are shown in Fig. 10. The EDXA analysis depicted in Table 3 revealed that the corrosion products had high contents of oxygen and aluminum. According to the Raman spectra in Fig. 8c, the white particles on the surface of the TMAZ, BM, and NZ were $Na_2SO_4$. Cracks were also observed in the HAZ (Fig. 10c) and formed as a result of cathodic reduction [4]. Cathodic reduction on the constituent particles can increase the alkalinity in the surrounding solution, leading to the dissolution of the aluminum matrix. SEM/EDXA results agree well with those from the Raman analysis (Fig. 8c).

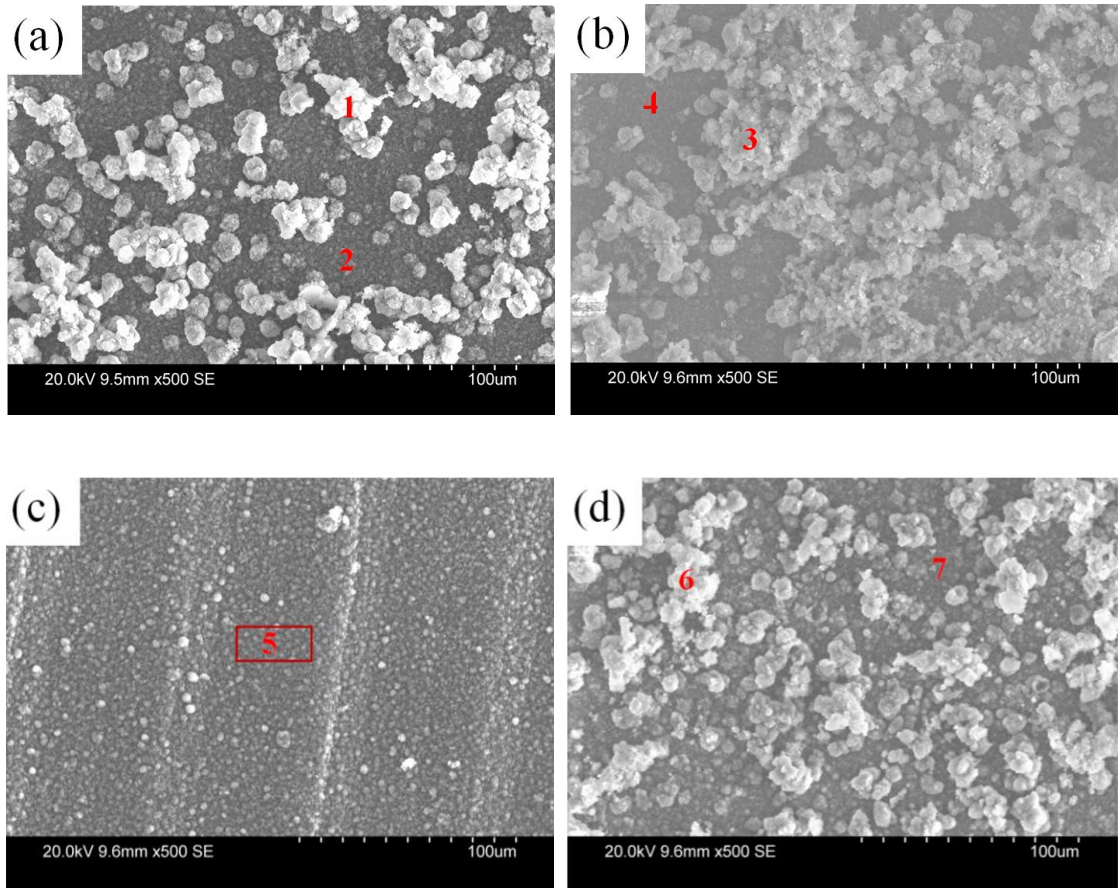

Fig. 11 SEM images of FSW AA6061-AA6061 after 90 days of immersion in 3.15 wt% NaCl: (a) NZ, (b) TMAZ, (c) HAZ, and (d) BM

Table 4 EDXA quantification results of eight points in Fig. 11

| Element [at. %] | C | O | Na | Al | Cl |
|---|---|---|---|---|---|
| Spectrum 1 (NZ) | 0.00 | 78.02 | 1.23 | 19.46 | 1.29 |
| Spectrum 2 (NZ) | 0.00 | 77.57 | 0.00 | 21.63 | 0.80 |
| Spectrum 3 (TMAZ) | 0.00 | 69.09 | 9.03 | 14.32 | 7.56 |
| Spectrum 4 (TMAZ) | 0.00 | 79.47 | 0.00 | 19.94 | 0.60 |
| Spectrum 5 (HAZ) | 0.00 | 77.34 | 0.00 | 22.35 | 0.31 |
| Spectrum 6 (BM) | 3.59 | 78.01 | 0.00 | 18.04 | 0.36 |
| Spectrum 7 (BM) | 0.00 | 76.17 | 0.37 | 22.32 | 1.14 |

Fig. 11 shows the SEM images of four regions (NZ, TMAZ, HAZ, and BM) of FSW 6061-6061 after 90 days of immersion in 3.15 wt% NaCl. EDXA quantification results shown in Table 4 reveal high concentrations of oxygen and aluminum, which agrees well with the Raman results (Fig. 9 c). The main corrosion product on the four regions was Al(OH)$_3$. As shown in Fig. 11, there was more corrosion product in the TMAZ, BM, and NZ than in the HAZ, which means that the corrosion resistance of the HAZ was better than the TMAZ, BM, and NZ. The low corrosion on the NZ could be caused by the microstructure or the specimen orientation.

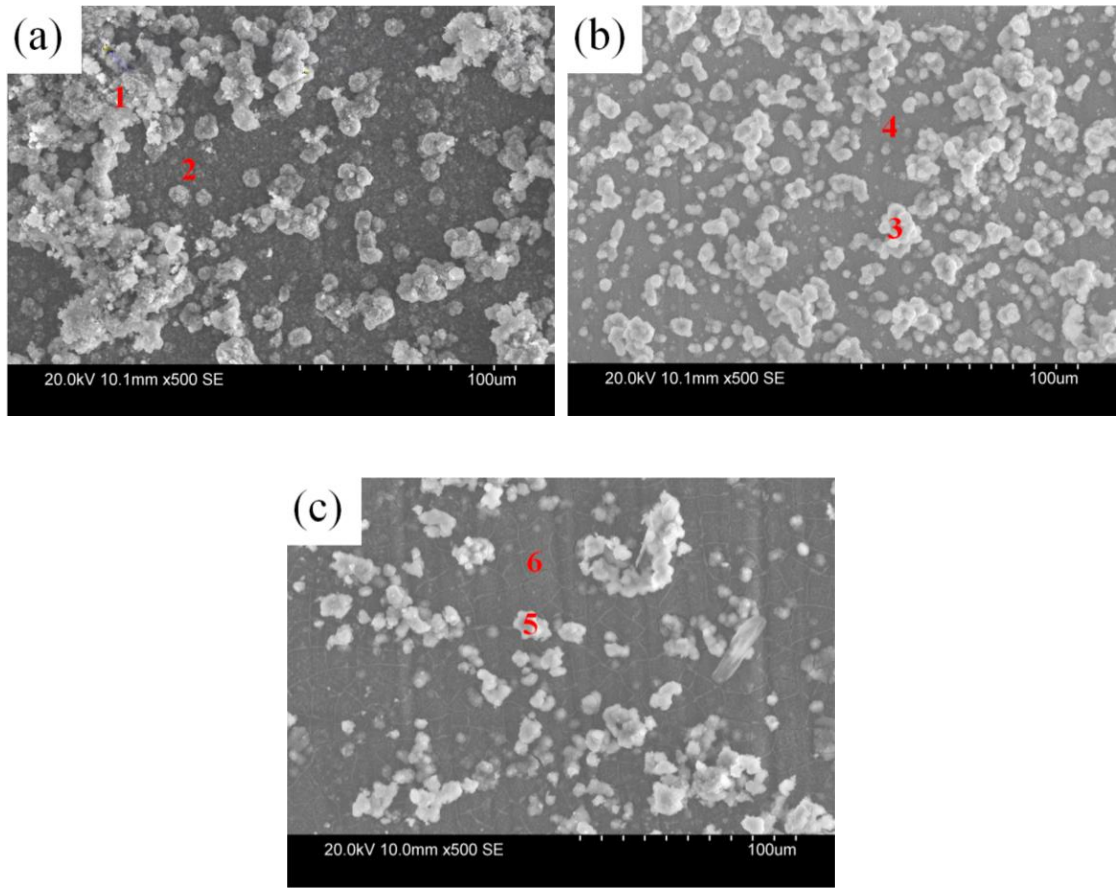

Fig. 12 SEM images of NZ of FSW AA6061-AA6061 after 120 days immersion in (a) 3.15 wt% NaCl, (b) ASTM seawater, and (c) 0.5 M Na2SO4 solution

Table 5 EDXA quantification results of six points in Fig. 12

| Element [at. %] | C | S | O | Na | Al | Cl | Mg |
| --- | --- | --- | --- | --- | --- | --- | --- |
| Spectrum 1 | 0.00 | 0.00 | 76.50 | 3.40 | 15.85 | 4.25 | 0.00 |
| Spectrum 2 | 0.00 | 0.00 | 73.26 | 0.31 | 25.19 | 1.24 | 0.00 |
| Spectrum 3 | 0.00 | 3.64 | 65.95 | 2.84 | 13.90 | 6.23 | 7.44 |
| Spectrum 4 | 7.00 | 2.03 | 59.69 | 1.43 | 23.91 | 2.66 | 3.28 |
| Spectrum 5 | 14.97 | 3.11 | 63.18 | 6.00 | 12.74 | 0.00 | 0.00 |
| Spectrum 6 | 0.00 | 3.08 | 58.48 | 3.94 | 34.50 | 0.00 | 0.00 |

Fig. 12 showed SEM of NZ of FSW AA6061-AA6061 after 120 days immersion in 3.15 wt% NaCl, ASTM seawater, and 0.5 M $Na_2SO_4$ solution. According to Raman analysis (Fig. 9), the corrosion products on the surface of the coupon immersed in 3.15 wt% NaCl and ASTM seawater was $Al(OH)_3$. A denser layer of corrosion products appeared to cover the coupon exposed in the NaCl solution (Fig. 12a) as compared to the coupon exposed in ASTM seawater (Fig. 12b). The oxygen concentration of the two locations shown in Fig. 12a (NaCl solution) was also higher than that in Fig. 12b (ASTM seawater. As shown in Fig. 12c, cracking on the surface of the specimen exposed to the 0.5 M $Na_2SO_4$ solution was also found. The cracks are likely to be in the aluminum oxide corrosion layer.

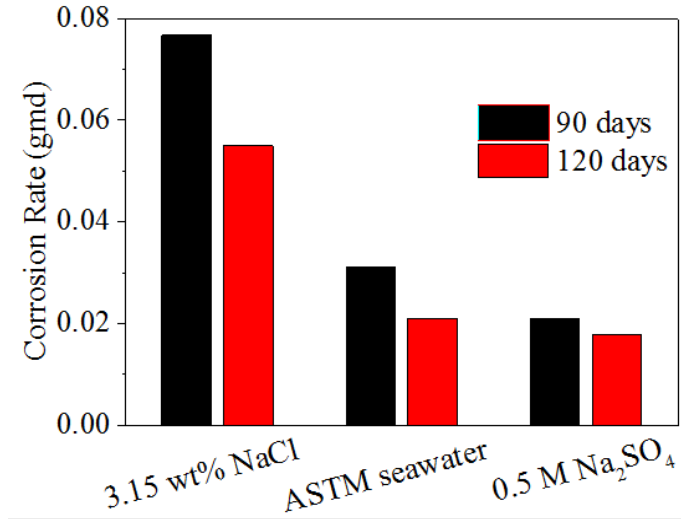

Fig. 13 Corrosion Rate of specimens after 90 and 120 days immersion in 3.15 wt% NaCl, ASTM seawater, and 0.5 M Na2SO4 solution for FSW AA6061-AA6061

The results of corrosion rate FSW AA6061-AA6061 by weight measurement were shown in Fig. 13. The corrosion rate of FSW AA6061-AA6061 decreased from 90 days to 120 days in three solutions. The corrosion rate from highest to lowest for three solutions was as follows: 3.15 wt% NaCl > ASTM seawater > 0.5 M $Na_2SO_4$. The maximum value of corrosion rate was obtained in 3.15 wt% NaCl solution after 90 days immersion.

*4 Conclusions*

An approach of the microstructure, mechanical properties, and corrosion behavior of FSW AA6061-AA6061 aluminum alloys had been made. The order of average grain size in different weld zones was as follow: BM > HAZ > TMAZ > NZ. The minimum hardness of 85.0 HV was obtained in the HAZ region, and the maximum value of 106.5 HV was present in the BM. The tensile and yield strengths of the weld zones were less than that of the BM tensile specimens. In comparisons to the properties of the BM, the ductility increased in the longitudinal tensile test specimens (that consisted of the NZ), but decreased in the transverse tensile specimens that cut through all of the weld zones. Fracture occurred in the HAZ region, which had the lowest hardness of all of the weld zones. The friction stir welding improved the corrosion resistance and the HAZ had better corrosion resistance than other regions. Raman results revealed $Al(OH)_3$ as the main corrosion product on coupons immersed in the three solutions. The oxygen concentration of the various weld zones increased, but the aluminum concentration decreased as immersion time increased.